\documentclass[twocolumn,showpacs,pra]{revtex4}
\usepackage{amsmath,amssymb}
\usepackage{graphicx}
\usepackage{verbatim}

\begin{document}

\title{Relativistic quantum theory of high harmonic generation on atoms/ions
by strong laser fields}
\author{H. K. Avetissian}
\email{avetissian@ysu.am}
\author{A. G. Markossian}
\author{G. F. Mkrtchian}
\affiliation{Centre of Strong Fields Physics, Yerevan State University, 1 A. Manukian,
Yerevan 0025, Armenia }

\begin{abstract}
High-order harmonic generation (HHG) by hydrogenlike atoms/ions in the
uniform periodic electric field, formed by the two linearly polarized
counterpropagating laser beams of relativistic intensities, is studied. The
relativistic quantum theory of HHG in such fields, at which the impeding
factor of relativistic magnetic drift of a strong wave can be eliminated, is
presented arising from the Dirac equation. Specifically, a scheme of HHG in
underdense plasma with the copropagating ultraintense laser and fast ion
beams is proposed.
\end{abstract}

\pacs{42.65.Ky, 32.80.Rm, 31.30.J-, 03.65.Pm}
\date{\today }
\maketitle

\section{Introduction}

The unprecedented progress of laser technology in the recent decade made
available the realization of femtosecond laser sources with relativistic: $%
10^{18}\ \mathrm{W/cm}^{2}$ (in optical domain) and ultrarelativistic
intensities up to $10^{22}\ \mathrm{W/cm}^{2}$\ by the chirped pulse
amplification technique \cite{Chirp}. It is expected that the next
generation of ultrarelativistic lasers will significantly exceed these
intensities in the near future \cite{ELI}. The interaction of such powerful
electromagnetic radiation with matter opens real possibilities for
revelation of ultrafast laser-matter interaction dynamics in supershort time
scales with new research fields, such as Relativistic Optics \cite%
{Mourou,Rel-Rep1,Rel-Rep2} and Attoscience \cite{Atto1,Atto2,Atto3}, as well
as many nonlinear electrodynamic effects at the ultrarelativistic
interaction with the QED vacuum and accelerator beams \cite{Avet-Book}.%
\textrm{\ }

At the laser-matter interaction the high-order harmonic generation (HHG) by
atoms/ions \cite{Main1,Main2,Main3,Main4} is a subject of extreme interest,
since HHG is one of the most promising mechanisms for creation of coherent
XUV sources \cite{Report-old,Report1} and intense attosecond pulses \cite%
{Atto3}, in particular, for the control of electrodynamic processes on
attosecond time scales. To reach a far X-ray region one needs the atoms or
ions with a large nuclear charge and laser fields of ultrahigh intensities
at which the nondipole interaction and relativistic effects become essential 
\cite{Relat1,Relat2,Relat3,Relat4}. At such intensities the state of ionized
electron becomes relativistic already at the distances $l\ll \lambda $ ($%
\lambda $ is the wavelength of a laser radiation) irrespective of its
initial state, hence the investigation of laser-atom/ion induced processes
such as Stimulated Bremsstrahlung (SB), Above Threshold Ionization (ATI)
etc., require relativistic consideration \cite{mer1,mer2,mer3}.
Specifically, the relativistic drift of a photoelectron due to the magnetic
field of a strong electromagnetic wave becomes the major inhibiting factor
in the relativistic regime of HHG. Due to this drift the significant HHG
suppression takes place and diverse schemes have been proposed to overcome
this negative effect \cite{scheme4,scheme5,scheme3}. For compensation of the
magnetic drift in relativistic domain, in Refs. \cite{scheme4}, \cite%
{scheme5} two counterpropagating laser pulses of linear and circular
polarizations were considered, respectively, at which the configuration of a
standing wave is formed. The effect of resulting magnetic field of the
standing wave is vanished near the stationary maxima at the waves' linear
polarizations \cite{scheme4}. While at the circular polarizations of
counterpropagating laser pulses one can achieve the fully vanishing of
resulting longitudinal magnetic force, responsible for magnetic drift \cite%
{scheme5}. However, the latter takes place at the adiabatic turn on of the
waves \cite{Avet-Book}, which is not valid for short laser pulses of
relativistic intensities.

So, a standing wave configuration formed by the two counterpropagating laser
beams of linear polarizations is of interest due to the simplicity to
realize such field structure providing incomparable large HHG rates in the
relativistic regime. At the lengths much smaller than a wavelength of a pump
wave the effective field of the standing wave may be approximated by the
uniform periodic electric field.\ Note that the consideration of the HHG in
the field of a standing wave in the paper \cite{scheme4} is based on the
semiclassical model, and only the first order relativistic effects have been
taken into account. The entire relativistic quantum-mechanical consideration
of the HHG in a standing wave or in a nonstationary strong electric field on
the base of the Dirac equation is still lacking.

To overcome the suppression of the HHG in the relativistic regime caused by
the magnetic field of a strong laser pulse, in the present work the
relativistic HHG by hydrogenlike ions in the uniform periodic electric
field, formed by the two linearly polarized counterpropagating laser beams
of relativistic intensities, at the distances much less than a laser
wavelength, is studied. We propose also a scheme of relativistic HHG in
plasma, where a traveling laser pulse in the own frame of reference (moving
with the pulse group velocity in plasma) is transformed into the uniform
periodic electric field, i.e. the wave magnetic field in this case vanishes
completely, in contrast to a standing wave configuration formed by the
counterpropagating laser beams. So, it is also feasible the effective HHG in
plasma with the copropagating ultraintense laser and fast ion beams of the
same velocities. Note that in Refs. \cite{scheme1,scheme1.2} a scheme with
counterpropagating relativistic ion and laser beams was considered, where
due to the large Doppler upshift the high harmonic frequency may appear in
the $10\ \mathrm{keV}$-$1\ \mathrm{MeV}$ domain.

The organization of the paper is as follows. In Sec. II we present analytic
and fully relativistic quantum theory of HHG arising from the Dirac
equation. In Sec. III with the help of Fast Fourier Transform algorithm we
present HHG spectrum for diverse ion and field parameters. Finally,
conclusions are given in Sec.\ IV.

\section{Basic model and theory}

Let two linearly polarized plane electromagnetic waves with carrier
frequency $\omega $ and amplitude of the electric field $\mathbf{E}_{a}$%
\begin{equation}
\mathbf{E}_{1}=\mathbf{E}_{a}\cos (\omega t-\mathbf{kr}),\quad \mathbf{E}%
_{2}=\mathbf{E}_{a}\cos (\omega t+\mathbf{kr}),  \label{waves}
\end{equation}%
propagating in the opposite directions in vacuum, interplay with the
hydrogenlike ions having the charge number of the nucleus $Z_{a}$. We will
assume that $\lambda >>a$, where $a$ is the characteristic size of the
atomic system and $\lambda $ is the wavelength of a pump wave (for the HHG
this condition is always satisfied).

At the photoionization of an atom/ion in the strong traveling wave field
taking into account the relativistic drift due to the magnetic field, one
can expect that the probability of ionized electron recombination with ionic
core could be non-negligible only if the electron is initially born with a
nonzero velocity oppositely directed to incident laser beam. The probability
of tunneling ionization with nonzero initial velocity is given by quantum
mechanical tunneling theory \cite{Delone} as: 
\begin{equation}
W_{ion}\varpropto e^{-2\left( Z_{a}^{2}+\mathrm{v}^{2}\right) ^{3/2}/(3E)},
\label{ion}
\end{equation}%
where $\mathrm{v}$ is the initial velocity, and $E$ is the electric field
strength of the wave (here and below, unless stated otherwise, we employ
atomic units). However, according to (\ref{ion}) the ionization probability
falls off exponentially if this velocity $\mathrm{v}$ becomes larger than
the characteristic atomic velocities, irrespective of it's direction. Since
we study the case of superstrong laser fields with $\xi \equiv E/c\omega
\sim 1$ ($\xi $ is the relativistic invariant parameter of the wave
intensity) when the energy of the interaction of an electron with the field
over a wavelength becomes comparable to the electron rest energy, the
required velocities becomes comparable with the light speed $c$ ($c=137$ $%
\mathrm{a.u.}$). Hence, the probability (\ref{ion}) in the such fields is
practically zero. Therefore, in considering case of a standing wave formed
by the laser beams (\ref{waves}), a significant input in the HHG process
will be conditioned by the ions situated near the stationary maxima of the
standing wave. For this points the magnetic fields of the counterpropagating
waves cancel each other. Since the HHG is essentially produced at the
lengths $l<<\lambda $, we will assume the effective field to be: 
\begin{equation}
\mathbf{E}(t)=\widehat{\mathbf{e}}E_{0}\cos \omega t,  \label{Et}
\end{equation}%
where $E_{0}=2E_{a}$, and $\widehat{\mathbf{e}}$ is the unit polarization
vector.

As is clear from the above consideration for HHG, one needs to exclude the
relativistic drift of a photoelectron due to the magnetic field of a
traveling wave, i.e. the magnetic component of the wave. This can also be
achieved in the plasma- like medium with a refractive index $n_{p}(\omega
)<1 $. Indeed, let a plane, transverse, and linearly polarized
electromagnetic wave with a frequency $\omega $ propagates in plasma ($%
\omega >\omega _{p}$; $\omega _{p}=\sqrt{4\pi N_{e}}$ is the plasma
frequency). For the electric and magnetic field strengths we have%
\begin{equation}
\mathbf{E}=\mathbf{E}_{a}\cos \left( \omega t-\mathbf{kr}\right) ,\ \mathbf{%
H=}\frac{c}{\omega }\left[ \mathbf{k}\times \mathbf{E}\right] ,  \label{PEt}
\end{equation}%
where $\left\vert \mathbf{k}\right\vert =n_{p}(\omega )\omega /c$. Suppose
that an ion beam copropagates together a laser beam in such plasma with a
mean velocity $\mathrm{V}$ equal to a laser beam group velocity: $\mathrm{V}%
=cn_{p}(\omega )$. To find out the HHG rate it is convenient to solve the
problem in the center-of-mass ($C$) frame of ions. In this frame the wave
vector of the photons is $\mathbf{k}^{\prime }=0$. The traveling
electromagnetic wave is transformed in the $C$ frame into the field \cite%
{Avet-Book}:%
\begin{equation*}
\mathbf{H}^{\prime }\equiv 0,\ \mathbf{E}^{\prime }\left( t^{\prime }\right)
=\mathbf{E}_{a}^{\prime }\cos \omega ^{\prime }t^{\prime },
\end{equation*}%
\begin{equation}
\omega ^{\prime }=\omega \sqrt{1-n_{p}^{2}(\omega )}=\omega _{p}.
\label{n20}
\end{equation}%
Hence, in the frame of ions center-of-mass the problem of HHG in a plasma is
reduced to one in vacuum in the field of the same configuration (\ref{Et})
with $\mathbf{E}_{a}^{\prime }=\mathbf{E}_{a}\omega _{p}/\omega $. When $%
\omega >>\omega _{p}$, one needs relativistic ion beams. Concerning the ion
beams of necessary relativism $\gamma =\left( 1-\mathrm{V}^{2}/c^{2}\right)
^{-1/2}=\omega /\omega _{p}$, note that the relativistic ion beams in
arbitrary charge states, with the Lorentz factor up to about $30$ is
supposed to be realized at the new accelerator complex at Gesellschaft f\"{u}%
r Schwerionenforschung (GSI) (Darmstadt,Germany). It is worthy to note that
if a laser frequency $\omega $ is close enough to plasma frequency $\omega
_{p}$, one can achieve the implementation of proposed scheme for HHG with
nonrelativistic ion beams. In this case one can use quasimonoenergetic and
low emittance ions bunches of solid densities generated from ultrathin foils
-nanotargets by supershort laser pulses of relativistic intensities \cite%
{Target}.

To find out the relativistic probabilities of HHG in the field (\ref{Et}) we
arise from the Dirac equation:%
\begin{equation}
i\frac{\partial \left\vert \Psi \right\rangle }{\partial t}=\left( c\widehat{%
\mathbf{\alpha }}\widehat{\mathbf{p}}+\widehat{\beta }c^{2}+\frac{Z_{a}}{r}-%
\mathbf{rE}\left( t\right) \right) \left\vert \Psi \right\rangle ,
\label{DE}
\end{equation}%
where $\widehat{\mathbf{\alpha }}$ and $\widehat{\beta }$ are the Dirac
matrices in the standard representation, $\mathbf{\sigma }=\left( \sigma
_{x},\sigma _{y},\sigma _{z}\right) $ are the Pauli matrices and $\mathbf{%
\hat{p}}$ is the operator of the kinetic momentum ($\widehat{\mathbf{p}}=-i%
\mathbf{\nabla }$). Without loss of generality one can take the polarization
vector $\widehat{\mathbf{e}}$ aligned with the $z$\ axis of spherical
coordinates.

We denote the atomic bound states by $\left\vert \eta \right\rangle $, where 
$\eta $ indicates the set of quantum numbers that characterizes the state $%
\eta =\{n,j,l,M\}$. Here $n$ is the principal quantum number, $j$ is the
whole moment, $l$ is the orbital moment and $M$ is the magnetic quantum
number.

Following the ansatz developed in the Ref. \cite{Main3}, the time dependent
wave function can be expanded as%
\begin{equation}
\left\vert \Psi \right\rangle =\left( C_{0}\left( t\right) \left\vert \eta
_{0}\right\rangle +\sum_{\mathbf{\mu }}\int d\mathbf{p}C_{\mathbf{\mu }%
}\left( \mathbf{p,}t\right) \left\vert \mathbf{p,}\mu \right\rangle \right)
e^{-i\varepsilon t}.  \label{Exp}
\end{equation}%
Here%
\begin{equation}
\left\vert \mathbf{p,}\mu \right\rangle =\frac{1}{\left( 2\pi \right) ^{3/2}}%
\sqrt{\frac{\mathcal{E}\left( \mathbf{p}\right) +c^{2}}{2\mathcal{E}\left( 
\mathbf{p}\right) }}%
\begin{pmatrix}
\varphi _{\mathbf{\mu }} \\ 
\frac{c\left( \mathbf{\sigma p}\right) }{\mathcal{E}\left( \mathbf{p}\right)
+c^{2}}\varphi _{\mathbf{\mu }}%
\end{pmatrix}%
e^{i\mathbf{pr}},  \label{Df}
\end{equation}%
are the Dirac free solutions \cite{Landau} with energy $\mathcal{E}\left( 
\mathbf{p}\right) =\sqrt{c^{2}\mathbf{p}^{2}+c^{4}}$ and polarization states 
$\mu =1,-1$:%
\begin{equation}
\varphi _{1}=%
\begin{pmatrix}
1 \\ 
0%
\end{pmatrix}%
,\qquad \varphi _{-1}=%
\begin{pmatrix}
0 \\ 
1%
\end{pmatrix}%
.  \label{Pb}
\end{equation}%
As an initial bound state wave function $\left\vert \eta _{0}\right\rangle $
we assume the ground-state bispinor wave function \cite{Landau} for the
hydrogenlike ion with the quantum numbers $n=1$, $j=1/2$, $l=0$, and $M=1/2$:%
\begin{equation}
\left\vert \eta _{0}\right\rangle =\frac{Z_{a}^{3/2}}{\sqrt{\pi \Gamma
\left( 3-2\epsilon \right) }}%
\begin{pmatrix}
\sqrt{2-\epsilon } \\ 
0 \\ 
i\cos \theta \sqrt{\epsilon } \\ 
i\sin \theta e^{i\varphi }\sqrt{\epsilon }%
\end{pmatrix}%
\left( 2rZ_{a}\right) ^{-\epsilon }e^{-Z_{a}r}.  \label{DC}
\end{equation}%
Here $\Gamma \left( x\right) $ is the Euler gamma function, $\theta \ $and $%
\varphi $ are the polar and azimutal angles, $\epsilon =1-\sqrt{%
1-Z_{a}^{2}/c^{2}},$ and $\varepsilon =c^{2}\left( 1-\epsilon \right) $ is
the energy of the ground state. We assume $Z_{a}<137$, and the parameter $%
\epsilon $ can take values lying in the interval $0<\epsilon <1$. In the
expansion (\ref{Exp}) we have excluded the negative energy states, since the
input of the particle-antiparticle intermediate states will lead only to
small corrections to the processes considered. Neglecting the depletion of
the ground state $C_{0}\left( t\right) \simeq 1$ and the free-free
transitions due to Coulomb field, the Dirac equation for $C_{\mathbf{\mu }%
}\left( \mathbf{p,}t\right) $ reads as:%
\begin{equation*}
\frac{\partial C_{\mu }\left( \mathbf{p,}t\right) }{\partial t}+E\left(
t\right) \frac{\partial C_{\mu }\left( \mathbf{p,}t\right) }{\partial p_{z}}%
+i\left( \mathcal{E}\left( \mathbf{p}\right) -\varepsilon \right) C_{\mu
}\left( \mathbf{p,}t\right)
\end{equation*}%
\begin{equation}
=i\mathcal{D}_{\mathbf{\mu }}\left( \mathbf{p}\right) E\left( t\right) +%
\frac{c^{2}E\left( t\right) \left( \mu p_{x}-ip_{y}\right) }{2\mathcal{E}%
\left( \mathbf{p}\right) \left( \mathcal{E}\left( \mathbf{p}\right)
+c^{2}\right) }C_{-\mu }\left( \mathbf{p,}t\right) ,  \label{C}
\end{equation}%
where%
\begin{equation}
\mathcal{D}_{\mathbf{\mu }}\left( \mathbf{p}\right) =\left\langle \mathbf{p,}%
\mu \right\vert z\left\vert \eta _{0}\right\rangle  \label{Dm}
\end{equation}%
is the atomic dipole matrix element for the bound-free transition. The
latter can be calculated with the help of integral \cite{Podolsky}:%
\begin{eqnarray*}
&&\int_{0}^{\pi }e^{ic_{1}\cos \theta \cos \Theta }J_{m}(c_{1}\sin \theta
\sin \Theta )P_{l}^{m}\left( \cos \theta \right) \sin \theta d\theta \\
&=&\left( \frac{2\pi }{c_{1}}\right) ^{1/2}i^{l-m}P_{l}^{m}\left( \cos
\Theta \right) J_{l+1/2}(c_{1}),
\end{eqnarray*}%
where $P_{l}^{m}\left( \cos \Theta \right) $ is an associated Legendre
function of degree $l$ and order $m$, $J_{m}$ is a Bessel function of order $%
m$. Thus, for $\mathcal{D}_{\mathbf{\pm 1}}\left( \mathbf{p}\right) $ we
have:%
\begin{widetext}
\begin{equation*}
\mathcal{D}_{\mathbf{1}}\left( \mathbf{p}\right) =i\frac{2^{3-\epsilon }}{%
\pi }\sqrt{\frac{\mathcal{E}+c^{2}}{2\mathcal{E}}}\frac{Z_{a}^{5/2-\epsilon
}\Gamma \left( 3-\epsilon \right) }{\sqrt{\Gamma \left( 3-2\epsilon \right) }%
}\frac{p_{z}}{\left( p^{2}+Z_{a}^{2}\right) ^{3-\epsilon }}
\end{equation*}%
\begin{equation*}
\times \left[ \frac{1}{4c\sqrt{2\epsilon }}\frac{Z_{a}^{3-\epsilon }}{2p^{2}}%
\left( \Upsilon ^{3-\epsilon }+\Upsilon ^{\dag 3-\epsilon }-\frac{i\left(
p^{2}+Z_{a}^{2}\right) }{\left( 2-\epsilon \right) Z_{a}p}\left( \Upsilon
^{2-\epsilon }-\Upsilon ^{\dag 2-\epsilon }\right) \right) +\frac{1}{%
\mathcal{E}+c^{2}}\sqrt{\frac{1}{2\left( 2-\epsilon \right) }}\frac{%
iZ_{a}^{3-\epsilon }}{8p}\right.
\end{equation*}%
\begin{equation}
\left. \times \left( \left( \Upsilon ^{3-\epsilon }-\Upsilon ^{\dag
3-\epsilon }\right) -\frac{2i\left( p^{2}+Z_{a}^{2}\right) }{\left(
2-\epsilon \right) Z_{a}p}\left( \Upsilon ^{2-\epsilon }+\Upsilon ^{\dag
2-\epsilon }\right) -\frac{2\Gamma \left( 1-\epsilon \right) }{\Gamma \left(
3-\epsilon \right) }\frac{\left( p^{2}+Z_{a}^{2}\right) ^{2}}{Z_{a}^{2}p^{2}}%
\left( \Upsilon ^{1-\epsilon }-\Upsilon ^{\dag 1-\epsilon }\right) \right) %
\right] ,  \label{D1}
\end{equation}%
\begin{equation}
\mathcal{D}_{-\mathbf{1}}\left( \mathbf{p}\right) =\frac{i\sqrt{\epsilon }%
Z_{a}^{5/2-2\epsilon }c\left( p_{x}+ip_{y}\right) }{2^{1+\epsilon }\pi p^{3}%
\sqrt{\Gamma \left( 3-2\epsilon \right) \mathcal{E}\left( \mathcal{E}%
+c^{2}\right) }}\left[ \frac{\Gamma \left( 1-\epsilon \right) }{\left(
p^{2}+Z_{a}^{2}\right) ^{1-\epsilon }}i\left( \Upsilon ^{1-\epsilon
}-\Upsilon ^{\dag 1-\epsilon }\right) -\frac{Z_{a}p\Gamma \left( 2-\epsilon
\right) }{\left( p^{2}+Z_{a}^{2}\right) ^{2-\epsilon }}\left( \Upsilon
^{2-\epsilon }+\Upsilon ^{\dag 2-\epsilon }\right) \right] ,  \label{D_1}
\end{equation}%
\end{widetext} where%
\begin{equation}
\Upsilon =1-i\frac{p}{Z_{a}},\ \Upsilon ^{\dag }=1+i\frac{p}{Z_{a}}.
\label{par}
\end{equation}

The last term in Eq. (\ref{C}) describes spin flip due to free-free
transitions and proportional to transverse momentum $p_{x,y}$. As will be
seen below, the main contribution to the HHG process is conditioned by the
electrons with momentum components $p_{x,y}=0$. Hence, we will neglect the
last term in Eq. (\ref{C}), which is valid for the fields $E<<E_{S}$, where $%
E_{S}=m^{2}c^{3}/e\hbar $ is the Schwinger field strength. Then, from Eq. (%
\ref{C}) for the probability amplitudes $C_{\mathbf{\mu }}\left( \mathbf{p,}%
t\right) $ we obtain:%
\begin{equation*}
C_{\mathbf{\mu }}\left( \mathbf{p,}t\right) =i\int_{0}^{t}dt^{\prime }%
\mathcal{D}_{\mathbf{\mu }}\left( \mathbf{p}+\frac{1}{c}\left( \mathbf{A}%
\left( t\right) \mathcal{-}\mathbf{A}\left( t^{\prime }\right) \right)
\right) E\left( t^{\prime }\right)
\end{equation*}%
\begin{equation}
\times \exp \left\{ -i\int_{t^{\prime }}^{t}\left[ \mathcal{E}\left( \mathbf{%
p+}\frac{1}{c}\left( \mathbf{A}\left( t\right) \mathbf{-A}\left( t^{\prime
\prime }\right) \right) \right) -\varepsilon \right] dt^{\prime \prime
}\right\} ,  \label{Cp}
\end{equation}%
where $\mathbf{A}\left( t\right) =-\widehat{\mathbf{e}}c^{2}\xi \sin \omega
t $ is the vector potential of resulting electric field.

For the harmonic radiation perpendicular to the polarization direction $%
\widehat{\mathbf{e}}$ one needs mean value of the $z$ component of the
electron current density $J\left( t\right) =c\left\langle \Psi \right\vert 
\widehat{\alpha }_{z}\left\vert \Psi \right\rangle $. Using Eqs. (\ref{Exp}%
), (\ref{Cp}) and neglecting the contribution by free-free transitions, we
obtain: 
\begin{equation*}
J\left( t\right) =i\sum_{\mathbf{\mu }}\int d\mathbf{p}\int_{0}^{t}dt^{%
\prime }\mathcal{J}_{\mu }\left( \mathbf{p}-\frac{1}{c}\mathbf{A}\left(
t\right) \right)
\end{equation*}%
\begin{equation*}
\times \mathcal{D}_{\mathbf{\mu }}\left( \mathbf{p-}\frac{1}{c}\mathbf{A}%
\left( t^{\prime }\right) \right) E\left( t^{\prime }\right)
\end{equation*}%
\begin{equation}
\times \exp \left\{ -iS\left( \mathbf{p},t,t^{\prime }\right) +i\varepsilon
\left( t-t^{\prime }\right) \right\} +\mathrm{c.c.},  \label{jz}
\end{equation}%
where 
\begin{equation*}
S(\mathbf{p,}t,t^{\prime })=\int_{t^{\prime }}^{t}\mathcal{E}\left( \mathbf{%
p-}\frac{1}{c}\mathbf{A}\left( t^{\prime \prime }\right) \right) dt^{\prime
\prime }
\end{equation*}%
\begin{equation}
=\int_{t^{\prime }}^{t}\sqrt{c^{2}\left( \mathbf{p+}\widehat{\mathbf{%
\epsilon }}c\xi \sin \omega t^{\prime \prime }\right) ^{2}+c^{4}}dt^{\prime
\prime }  \label{Sp}
\end{equation}%
is the relativistic classical action of an electron in the field and $%
\mathcal{J}_{\mu }\left( \mathbf{p}\right) =c\left\langle \eta
_{0}\right\vert \widehat{\alpha }_{z}\left\vert \mathbf{p,}\mu \right\rangle 
$.

As in the nonrelativistic case, the HHG rate is mainly determined by the
exponential in the integrand of Eq. (\ref{jz}) with exact relativistic
classical action. The integral over the intermediate momentum $\mathbf{p}$
and time $t^{\prime }$ can be calculated using saddle-point method. The
saddle momentum\ is determined by the equation:%
\begin{equation}
\frac{\partial S(\mathbf{p,}t,t^{\prime })}{\partial \mathbf{p}}=0,
\label{sad_p}
\end{equation}%
which for momentum components gives $p_{x,y}=0$, and the $z$ component of
momentum ($p_{s}$) is given by the solution of the equation: 
\begin{equation}
\int_{t^{\prime }}^{t}\frac{p_{s}\mathbf{+}c\xi \sin \omega t^{\prime \prime
}}{\sqrt{\left( p_{s}\mathbf{+}c\xi \sin \omega t^{\prime \prime }\right)
^{2}+c^{2}}}dt^{\prime \prime }=0.  \label{sad_r}
\end{equation}%
In contrast to nonrelativistic \cite{Main3} and relativistic \cite{Relat2}
cases for a traveling wave, this equation can not be solved analytically and
requires numerical solution. Integrating the latter over $\mathbf{p}$, for
the current density we obtain:%
\begin{equation*}
J(t)=\left( 2\pi \right) ^{3/2}\sqrt{i}\int_{0}^{t}dt^{\prime }\frac{%
e^{-i\left( S(\mathbf{p}_{s}\mathbf{,}t,t^{\prime })-\varepsilon \left(
t-t^{\prime }\right) \right) }}{\sqrt{\left\vert \det S_{\mathbf{pp}%
}^{\prime \prime }\right\vert }}E\left( t^{\prime }\right)
\end{equation*}%
\begin{equation}
\times \mathcal{D}_{\mathbf{1}}\left( \mathbf{p}_{s}\mathcal{-}\frac{1}{c}%
\mathbf{A}\left( t^{\prime }\right) \right) \mathcal{J}_{1}\left( \mathbf{p}%
_{s}-\frac{1}{c}\mathbf{A}\left( t\right) \right) +\mathrm{c.c.,}
\label{jz2}
\end{equation}%
where%
\begin{eqnarray}
\det S_{\mathbf{pp}}^{\prime \prime } &=&\left( \int_{t^{\prime }}^{t}\frac{%
d\tau }{\gamma \left( t^{\prime },\tau ;\xi \right) }\right)
^{2}\int_{t^{\prime }}^{t}\frac{d\tau }{\gamma ^{3}\left( t^{\prime },\tau
;\xi \right) },  \notag \\
\gamma \left( t^{\prime },\tau ;\xi \right) &=&\left( 1+\left( p_{s}\mathbf{+%
}c\xi \sin \omega \tau \right) ^{2}/c^{2}\right) ^{1/2}.  \label{deter}
\end{eqnarray}%
Note that contribution from spin flip due to bound-free transitions in the
saddle-point vanishes: $\left. \mathcal{D}_{-\mathbf{1}}\right\vert
_{p_{x,y}=0}=0$. At $\xi <<1$, for $\det S_{\mathbf{pp}}^{\prime \prime }$
one can recover nonrelativistic result: $\det S_{\mathbf{pp}}^{\prime \prime
}=\left( t-t^{\prime }\right) ^{3}$.

The complex saddle times $t_{s}$ are the solutions of the following
equation: 
\begin{equation}
\frac{\partial S(\mathbf{p}_{s}\mathbf{,}t,t^{\prime })}{\partial t^{\prime }%
}+\varepsilon =0,  \label{sa_t}
\end{equation}%
which may be expressed by the transcendental equation 
\begin{equation}
\sqrt{c^{2}\left( p_{s}\mathbf{+}c\xi \sin \omega t_{s}\right) ^{2}+c^{4}}%
-\varepsilon =0.  \label{sa_t2}
\end{equation}%
Then expressing the saddle time as $t_{s}=t_{b}+i\delta $, with $\omega
\delta <<1$, one can obtain the saddle momentum 
\begin{equation}
p_{s}=-c\xi \sin \omega t_{b},  \label{ps}
\end{equation}%
and the imaginary part of the saddle time:%
\begin{equation}
\delta =\frac{Z_{a}}{\left\vert E\left( t_{b}\right) \right\vert },
\label{Imt}
\end{equation}%
where $E\left( t_{b}\right) =c\xi \omega \cos \omega t_{b}$. Taking into
account Eq. (\ref{ps}), from Eq. (\ref{sad_r}) for the real part of the
saddle time we obtain 
\begin{equation}
\int_{t_{b}}^{t}\frac{\sin \omega t^{\prime \prime }-\sin \omega t_{b}}{%
\sqrt{1+\xi ^{2}\left( \sin \omega t^{\prime \prime }-\sin \omega
t_{b}\right) ^{2}}}dt^{\prime \prime }=0.  \label{Ret}
\end{equation}%
As usual, $t_{b}$ is interpreted as the birth time of the photoelectron
which returns at the moment $t$ to the core and generates harmonic
radiation. The transition dipole moment has singularity at the saddle times
and the integration has been made with the help of the formula:%
\begin{eqnarray}
&&\int g(x)\frac{e^{-\lambda f\left( x\right) }}{\left( x-x_{0}\right) ^{\nu
}}dx\simeq i^{\nu }\sqrt{\pi }g(x_{0})\left[ 2f^{\prime \prime }\left(
x_{0}\right) \lambda \right] ^{\frac{\nu -1}{2}}  \notag \\
&&\times \frac{\Gamma \left( \nu /2\right) e^{-\lambda f\left( x_{0}\right) }%
}{\Gamma \left( \nu \right) }.  \label{asim}
\end{eqnarray}%
Thus, taking into account Eqs. (\ref{sad_p}-\ref{asim}), we obtain the
ultimate formula for current density:%
\begin{equation}
J(t)=\sum\limits_{t_{b}}C_{\mathrm{ion}}\left( t_{b}\right) C_{\mathrm{pr}%
}\left( t,t_{b}\right) C_{\mathrm{rec}}\left( t,t_{b}\right) +\mathrm{c.c.}.
\label{main}
\end{equation}

\begin{figure}[tbp]
\includegraphics{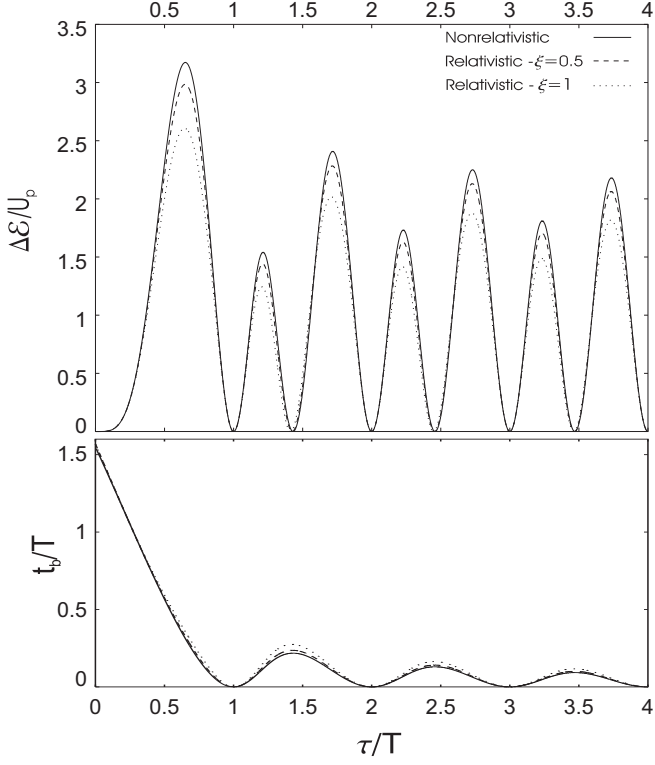}
\caption{The photoelectron energy gain in units of ponderomotive energy $%
U_{p}=c^{2}\protect\xi ^{2}/4$ and ionization time versus the electron time
evolution in the continuum (return time) for the various laser intensities. }
\label{eps1.11}
\end{figure}

The formula (\ref{main}) is the analogous to the nonrelativistic formula for
the dipole moment in the three step model \cite{Main4}. Here the summation
is carried out over the solutions of Eq. (\ref{Ret}). The tunneling
ionization amplitude $C_{\mathrm{ion}}\left( t_{b}\right) $ is: 
\begin{equation*}
C_{\mathrm{ion}}\left( t_{b}\right) =\frac{2i}{\sqrt{\pi }}\frac{\Gamma
\left( \frac{3-\epsilon }{2}\right) }{\sqrt{\Gamma \left( 3-2\epsilon
\right) }}\frac{\left\vert E\left( t_{b}\right) \right\vert ^{\frac{\epsilon 
}{2}}}{2^{\frac{3\epsilon }{2}}E\left( t_{b}\right) }
\end{equation*}%
\begin{equation}
\times \frac{Z_{a}^{\frac{3-3\epsilon }{2}}}{\left( 1-\epsilon \right) ^{%
\frac{1-\epsilon }{2}}}e^{-\frac{Z_{a}}{\left\vert E\left( t_{b}\right)
\right\vert }\left( c^{2}\epsilon -\frac{Z_{a}^{2}}{6}\right) }.
\label{Cion}
\end{equation}%
\ The propagation amplitude is given by the expression 
\begin{equation}
C_{\mathrm{pr}}\left( t,t_{b}\right) =\frac{\left( 2\pi \right) ^{3/2}}{%
\sqrt{i}}\frac{\exp \left\{ -iS\left( \mathbf{p}_{s}\mathbf{,}t,t_{b}\right)
+i\varepsilon \left( t-t_{b}\right) \right\} }{\sqrt{\left\vert \det S_{%
\mathbf{pp}}^{\prime \prime }\right\vert }},  \label{Cpr}
\end{equation}%
and the recombination amplitude is: $C_{\mathrm{rec}}\left( t,t_{b}\right) =%
\mathcal{J}_{1}\left( \mathbf{p}_{s}-\frac{1}{c}\mathbf{A}\left( t\right)
\right) $, where 
\begin{widetext}
\begin{equation*}
\bigskip \mathcal{J}_{1}\left( \mathbf{p}\right) =\frac{1}{\pi }\sqrt{\frac{%
\mathcal{E}+c^{2}}{2\mathcal{E}}}\frac{Z_{a}^{7/2-2\epsilon }2^{-\epsilon }}{%
\sqrt{2\Gamma \left( 3-2\epsilon \right) }}\frac{cp_{z}\Gamma \left(
2-\epsilon \right) }{\left( p^{2}+Z_{a}^{2}\right) ^{2-\epsilon }}
\end{equation*}%
\begin{equation}
\bigskip \mathcal{\times }\left[ i\frac{c\sqrt{\left( 2-\epsilon \right) }}{%
\left( \mathcal{E}+c^{2}\right) p}\left( \Upsilon ^{2-\epsilon }-\Upsilon
^{\dag 2-\epsilon }\right) -\frac{\sqrt{\epsilon }}{p^{2}}\left( \Upsilon
^{2-\epsilon }+\Upsilon ^{\dag 2-\epsilon }-i\frac{\Gamma \left( 1-\epsilon
\right) }{1-\epsilon }\frac{\left( p^{2}+Z_{a}^{2}\right) }{pZ_{a}}\left(
\Upsilon ^{1-\epsilon }-\Upsilon ^{\dag 1-\epsilon }\right) \right) \right] .
\label{Crec}
\end{equation}
\end{widetext}

\begin{figure}[tbp]
\includegraphics{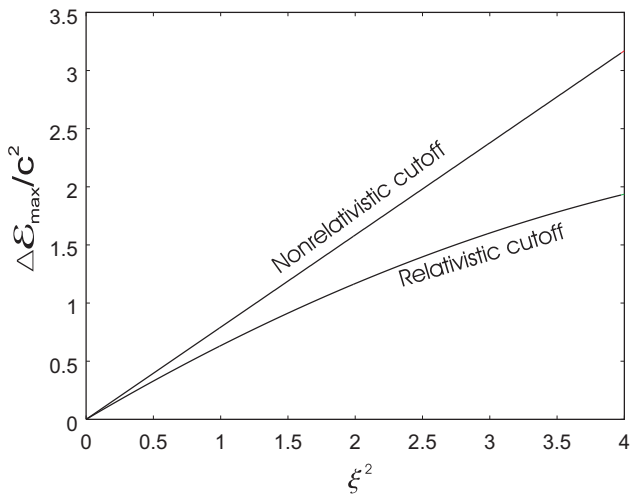}
\caption{Maximum energy gain of photoelectron, which defines the cutoff
frequency as a function of the relativistic parameter of the wave intensity.}
\label{eps1.119}
\end{figure}

\section{Relativistic Harmonic Spectra for Diverse Ion and Field Parameters}

In this section we present some numerical simulations for HHG when
relativistic effects become essential. In considering relativistic case the
saddle time and energy gain essentially depend on the intensity of the pump
wave. The latter leads to the modification of the HHG spectrum compared with
nonrelativistic one. In Fig. 1 we present the solution of Eq. (\ref{Ret})
for the born times $t_{b}$ which are limited to a quarter of the laser
period. The Fig. 1 also illustrated the photoelectron energy gain versus the
electron's time evolution in the continuum (return time) for the various
laser intensities. It is well known that the cutoff frequency of HHG is
defined by the maximal classical energy gain. For this reason in Fig. 2 the
maximum energy gain of the photoelectron as a function of the relativistic
parameter of the wave intensity $\xi ^{2}$ is displayed. As we see, the
relativistic cutoff essentially differs from the nonrelativistic one for $%
\xi \gtrsim 1$ and the shift of the cutoff position to the lower values of
the harmonic order for the same laser intensity becomes evident. The Fig. 1
also reveals the multiplateau character of the harmonic spectrum like to the
nonrelativistic one.

To find out harmonic spectrum, the Fast Fourier Transform algorithm has been
used. 
\begin{figure}[tbp]
\includegraphics{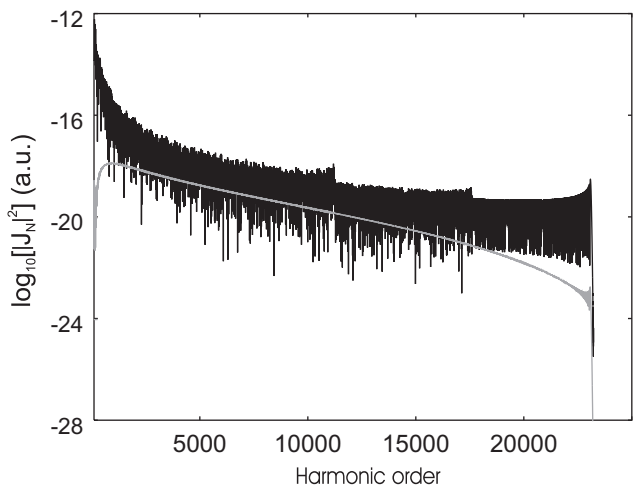}
\caption{Harmonic emission rate via $\log _{10}\left( \left\vert
J_{N}\right\vert ^{2}\right) $, as a function of the harmonic order, for an
ion with $Z_{a}=3$, $\protect\xi =0.3$, and frequency $\protect\omega %
=0.057\ \mathrm{a.u.}$ ($800$ $\mathrm{nm}$). The gray curve represents the
HHG spectrum for a traveling wave.}
\end{figure}
\begin{figure}[tbp]
\includegraphics{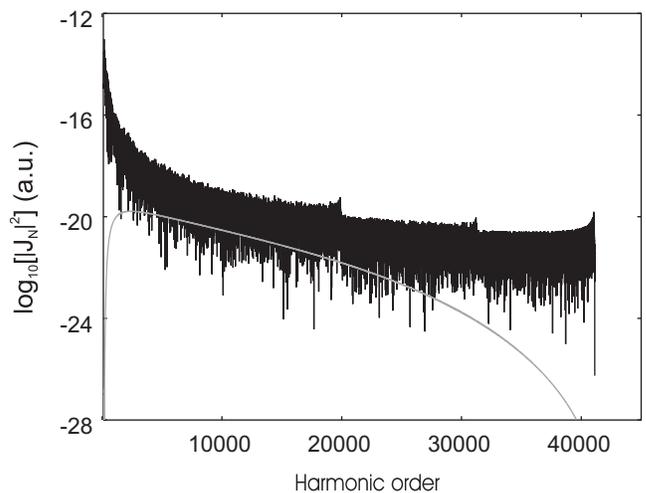}
\caption{The same, as in Fig. 3, but for the frequency $\protect\omega %
=0.043\ \mathrm{a.u.}$($1054$ $\mathrm{nm}$) and $\protect\xi =0.35$.}
\end{figure}

\begin{figure}[tbp]
\includegraphics{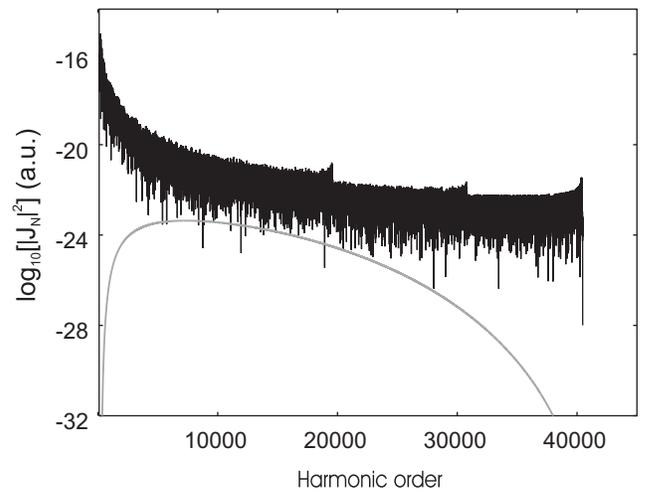}
\caption{Harmonic emission rate for an ion with $Z_{a}=4,$ standing wave
intensity $3.4\times 10^{17}\ \mathrm{W/cm}^{2}$ ($\protect\xi =0.4$), and
frequency $\protect\omega =0.057\ \mathrm{a.u.}$($800$ $\mathrm{nm}$). The
gray curve represents the HHG spectrum for a traveling wave.}
\label{eps1.5}
\end{figure}
Figures 3 and 4 display harmonic emission rate via modulus squared of the
Fourier transform of the current density $\left\vert J_{N}\right\vert ^{2}$
as a function of the harmonic order $N$ for an ion with $Z_{a}=3$. The
frequencies and relativistic parameters of intensity are: $\omega =0.057\ 
\mathrm{a.u.}$ ($\xi =0.3$) and $\omega =0.043\ \mathrm{a.u.}$ ($\xi =0.35$%
), respectively. Figure 5 displays harmonic emission rate for an ion with $%
Z_{a}=4$ and field intensity $3.4\times 10^{17}\ \mathrm{W/cm}^{2}$ ($\xi
=0.4$) with the wave frequency: $\omega =0.057\ \mathrm{a.u.}$. For the
comparison we have also brought the spectra for a traveling wave which was
obtained using the results of Ref. \cite{Relat3}. As we see from these
figures, with the increase of the laser intensity the HHG rate with a
standing wave field by many orders of magnitude is larger than HHG rate with
a traveling wave. Besides, the relativistic quantum mechanical harmonic
cutoff is shifted to the lower values compared with nonrelativistic one.

\begin{figure}[tbp]
\includegraphics{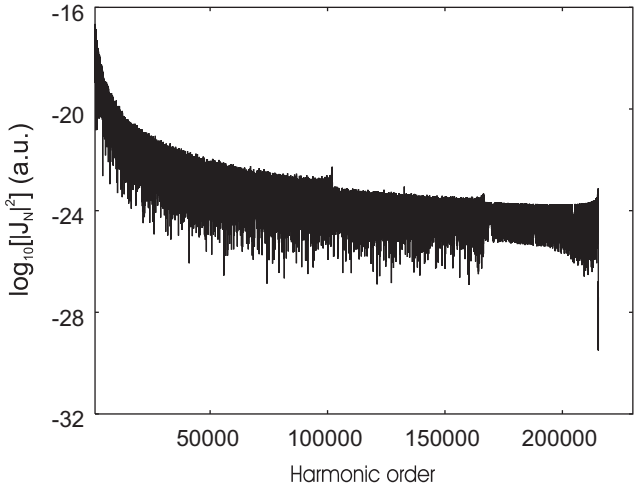}
\caption{Harmonic emission rate as a function of the harmonic order for an
ion with $Z_{a}=7$, $\protect\xi =1$, and $\protect\omega =0.184$ ($248$ $%
\mathrm{nm}$).}
\end{figure}

\begin{figure}[tbp]
\includegraphics{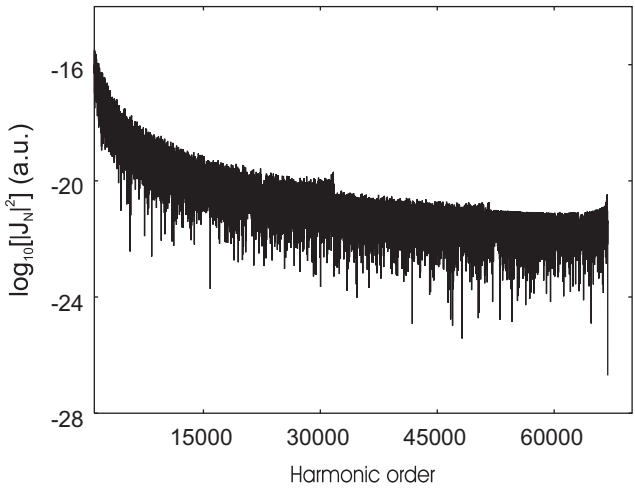}
\caption{Harmonic emission rate as a function of the harmonic order for an
ion with $Z_{a}=5$, $\protect\xi =1$, and $\protect\omega =0.057$ $\mathrm{%
a.u.}$.}
\end{figure}
\begin{figure}[tbp]
\includegraphics{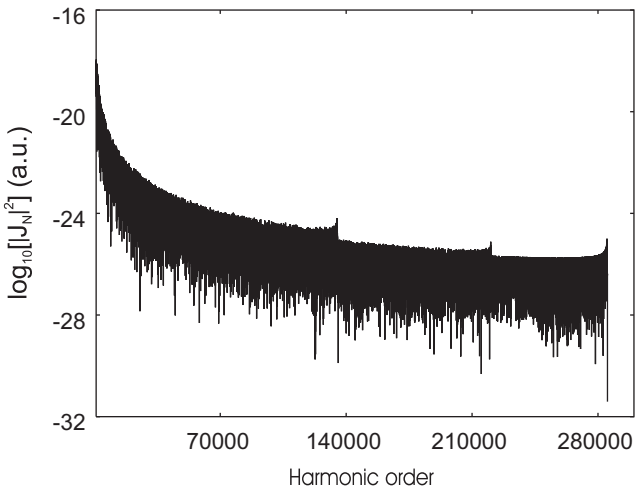}
\caption{The same, as in Fig. 7, but for the frequency $\protect\omega %
=0.043\ \mathrm{a.u.}$.}
\end{figure}

Figures 6, 7, and 8 display harmonic emission rates when the relativistic
parameter of the wave intensity is: $\xi =1$. In this case the relativistic
effects are essential. Figure 6 corresponds to $\omega =0.184\ \mathrm{a.u.}$
and $Z_{a}=7$. Figures 7 and 8 correspond to $Z_{a}=5$ with $\omega =0.057\ 
\mathrm{a.u.}$ and $\omega =0.043\ \mathrm{a.u.}$, respectively. For the
last three figures the cutoff position is $\simeq 2.61U_{p}$, which is in
good agreement with semiclassical analysis (see Fig. 2). For these
parameters the harmonic emission rates with a traveling wave are negligible,
meanwhile the relativistic rates with a standing wave field at photons
energies approaching to $\mathrm{MeV}$ region, are considerable.

\section{Conclusion}

We have presented a theoretical treatment of the HHG by hydrogenlike ions
with a large nuclear charge in the relativistic regime with a standing wave
configuration, formed by the two linearly polarized counterpropagating laser
beams of relativistic intensities, at the distances much less than a laser
wavelength. The investigation of the HHG with the such pump field
configuration is conditioned by known fact that in the relativistic regime
the significant HHG suppression takes place in the field of a strong
traveling wave because of the magnetic drift of a photoelectron along the
wave propagation direction. We have proposed a new scheme for relativistic
HHG with an ultraintense traveling laser pulse and copropagating ions in a
plasma. In this case the relativistic HHG occurs exactly without the
magnetic drift of the photoelectron, because the traveling wave in the ions
own frame of reference (moving with the laser pulse group velocity) in
underdense plasma is transformed into the purely uniform periodic electric
field (in contrast to a standing wave -- counterpropagating laser beams).
Hence, we expect that the efficient HHG with the current ultraintense laser
pulses and forthcoming relativistic ion beams (at GSI-Darmstadt accelerator)
will be feasible in the plasmas of conventional densities.

On the base of the Dirac equation we have presented the analytic and fully
relativistic quantum theory of HHG. With the help of the dynamic solution of
the Dirac equation and Fast Fourier Transform algorithm we have calculated
the harmonic spectrum. The obtained results have been applied to
hydrogenlike ions with a moderate nuclear charge. The harmonic spectrum
displays the significant difference compared both to non-relativistic and
relativistic spectra with a traveling laser pulse. In particular, the shift
of the cutoff position to the lower values of the harmonic order is
considerable. In the relativistic regime of interaction the harmonic
emission rates for a traveling laser pulse are negligible, meanwhile the
relativistic rates with a standing wave field, at photons energies
approaching to $\mathrm{MeV}$ region, are considerable.

\begin{acknowledgments}
This work was supported by SCS of RA under Project No. 10-3E-17
\end{acknowledgments}

\end{document}